\pdfoutput=1
\documentclass[10pt,a4paper,twocolumn]{paper}
\usepackage[utf8]{inputenc}
\usepackage{amsmath}
\usepackage{amsfonts}
\usepackage{amssymb}
\usepackage{graphicx}
\usepackage{nicefrac}
\usepackage{siunitx}
\usepackage{bm}
\usepackage{subfig}
\usepackage{url}
\usepackage[labelfont={bf,small},labelsep=space,format=plain,textfont={small}]{caption}
\usepackage{authblk}
\usepackage{upgreek}
\title{\vspace{-2.5cm} \center Thermodynamic Geodesics of a Reissner Nordstr\"{o}m Black Hole}
\author[1]{Christine R. Farrugia}
\author[1]{Joseph Sultana}
\affil[1]{Department of Mathematics, Faculty of Science, University of Malta, Msida MSD 2080, Malta}
\date{}                     
\begin{document}
  
\twocolumn[
\begin{@twocolumnfalse}
\maketitle
\begin{abstract}
Starting from a Geometrothermodynamics metric for the space of thermodynamic equilibrium states in the mass representation, we use numerical techniques to analyse the thermodynamic geodesics of a supermassive Reissner Nordstr\"{o}m black hole in isolation. Appropriate constraints are obtained by taking into account the processes of Hawking radiation and Schwinger pair--production. We model the black hole in line with the work of Hiscock and Weems \cite{Hiscock}. It can be deduced that the relation which the geodesics establish between the entropy $S$ and electric charge $Q$ of the black hole extremises changes in the black hole's mass. Indeed, the expression for the entropy of an extremal black hole is an exact solution to the geodesic equation. We also find that in certain cases, the geodesics describe the evolution brought about by the constant emission of Hawking radiation and charged-particle pairs.
\end{abstract}
\end{@twocolumnfalse}
]

\section{Introduction}
\label{intro}

Black holes are commonly thought of as regions of spacetime where gravity is so strong that it allows nothing, not even light, to escape. Such regions can be fully characterised by their mass, electric charge, and angular momentum, a property better known as the No--Hair Theorem of black holes \cite{Wheeler}. The black hole concept has its origins in the 18th century, when John Michell \cite{Michell} and Pierre Simon Laplace \cite{Gillispie} considered classical bodies with escape velocities exceeding the speed of light. Until the 1970s, black holes were thought of as `black', non-emitting objects at absolute zero. Things began to change when it became apparent that unless black holes were assigned an entropy, the second law of thermodynamics could be violated \cite{Bekenstein}. Bekenstein conjectured that this entropy would be proportional to the black hole area \cite{Bekenstein}. In 1973, Bardeen, Carter and Hawking put together a number of similarities between black hole mechanics and ordinary thermodynamics to formulate the four laws of black hole thermodynamics \cite{Bardeen}. The following year, Stephen Hawking discovered that black holes radiate particles continuously with a black body spectrum \cite{Hawking}; this radiation earned the name \emph{Hawking radiation}. Since then, black holes have increasingly been studied in terms of their thermodynamic properties. 

The last few decades have seen a growing interest in the use of geometry as a means of extracting important information about the thermodynamics of a system. The key points in the development of this practice are highlighted in \cite{Quevedo2007, Quevedo2011}. Stemming from the pioneering work of Gibbs \cite{Gibbs} and Carath\'{e}odory \cite{Caratheodory}, geometric thermodynamics refers to the modelling of thermodynamic systems in terms of differential manifolds. Riemannian geometry was introduced into thermodynamics by Rao in 1945 \cite{Rao}. In the 1970s, Hermann modelled the thermodynamic phase space as a manifold with contact structure \cite{Hermann}, while the first application of Riemannian geometry to the space of equilibrium states -- a subset of the phase space -- was due to Weinhold and Ruppeiner, who constructed metric structures on this space\footnote{To be precise, Weinhold worked in the tangent space defined at a general point of the equilibrium manifold, although it is possible to use his metric as a measure of distance in the manifold itself \cite{Andresen}} from the Hessian matrices of the internal energy \cite{Weinhold} and entropy \cite{Ruppeiner}, respectively. The two metric structures are conformally equivalent \cite{Salamon1984}. Using these metrics, Nulton et al \cite{Nulton} concluded that if a system undergoes a quasi-static thermodynamic process made up of $K$ steps, each equilibrating with a proper reservoir, the minimum changes in the availability and entropy of the Universe are proportional to the squared thermodynamic length of the path traversed. In other words, thermodynamic length controls the dissipation in finite-time processes. Indeed, starting from the Ruppeiner metric structure, it can be shown that the entropy produced irreversibly during a fixed thermodynamic time is least when the system evolves along a geodesic \cite{Diosi}. Crooks \cite{Crooks} considered how to define and computationally measure thermodynamic length for a small system described by equilibrium statistical mechanics. 

The formalism of Geometrothermodynamics (GTD) was put forward in recent years by Quevedo \cite{Quevedo2007}. In GTD, a system with $n$ thermodynamic degrees of freedom is described by a thermodynamic phase space, this being defined as a Riemannian contact manifold $(T', \Theta, G)$. $T'$ represents a $(2n+1)$-dimensional manifold equipped with a non-degenerate metric $G$, and $\Theta$ is a linear differential one-form with the property that\footnote{$\wedge$ stands for the exterior product, `$\text{d}$' the exterior derivative, and $(\text{d}\Theta)^n$ is equal to $\text{d}\Theta\wedge\dots\wedge \text{d}\Theta$, where $\text{d}\Theta$ appears $n$ times} $\Theta \wedge (\text{d}\Theta)^n \neq 0$. An $n$-dimensional submanifold $\varepsilon$ is defined by requiring that the smooth embedding map $\varphi:\varepsilon\rightarrow T'$ has a pullback $\varphi^*$ which satisfies $\varphi^*(\Theta)=0$; $\varepsilon$ is termed the \emph{space of thermodynamic equilibrium states} and its geometric properties, described by means of the metric $g=\varphi^*(G)$, yield information on the equilibrium thermodynamics of the corresponding physical system \cite{Quevedo2008}. 

The $(2n+1)$ coordinates of the phase space $T'$ consist of $n$ extensive variables $E^a$, $n$ conjugate intensive variables $I^a$ and the thermodynamic potential $\Phi$. The subset of extensive variables is usually chosen to coordinatize $\varepsilon$. The first law of thermodynamics, $\text{d}\Phi=I_b \text{d}E^b$, is satisfied on $\varepsilon$ -- and in turn, so are the conditions for equilibrium. In other words, $I_b=\partial \Phi/\partial E^b$ for all intensive variables $I^b$ \cite{Quevedo2008}. 

Unlike the Ruppeiner and Weinhold formalisms \cite{Salamon, Mrugala}, GTD is \emph{Legendre invariant} \cite{Quevedo2008}. Legendre transformations refer to the exchange of the role played by one or more extensive variables with that of the conjugate intensive ones, and invariance under such transformations ensures that the various thermodynamic potentials give rise to equivalent descriptions of the system \cite{Bravetti}. This is in line with equilibrium thermodynamics, in which the physical properties of a system are independent of the thermodynamic potential used to describe it \cite{Bravetti}.

As pointed out in \cite{Quevedo2008}, all the known field interactions have an associated curvature that acts as a measure of the interaction. This is also one of the benefits of the geometric description of thermodynamics embodied in GTD. More specifically, the curvature of the space of equilibrium states can serve to probe the thermodynamic interactions of the system -- for instance, curvature singularities indicate the presence of phase transitions. The link between geometry and thermodynamics provided by GTD has been investigated for a number of diverse systems, in works such as \cite{Quevedo2007, Quevedo2008, Quevedo2011, Bravetti2013, Quevedo2015, Alvarez, HernandoQuevedo, Bravetti2014, Han, AlessandroBravetti}. In particular, the geometrothermodynamics of the Reissner Nordstr\"om (RN) black hole are tackled in \cite{Bravetti2013, Alvarez, HernandoQuevedo}, and those of the asymptotically anti-de Sitter RN black hole in \cite{Quevedo2008}. Another point of interest is the new metric introduced in \cite{Hendi} to analyse the phase transition points of the heat capacity. This metric is partly based on GTD, and was used to study the geometric thermodynamics of charged black holes in Gauss--Bonnet--Massive Gravity \cite{Hendi2016} and in Brans--Dicke theory \cite{HendiSH2016}, among others. 

Equipped with GTD, it becomes possible to investigate the thermodynamic geodesics in the space of equilibrium states by extremising the thermodynamic `length' $\int \sqrt{g_{ab}\text{d}E^a\text{d}E^b}$. However, not all solutions to the geodesic equations are necessarily competent with the laws of thermodynamics. Those that do satisfy these laws represent quasi-static thermodynamic processes, which can hence be interpreted as a dense collection of equilibrium states (see `Thermodynamic systems as bosonic strings' by V\'{a}zquez, Quevedo and S\'{a}nchez [arXiv:0805.4819v5]). In `A geometric approach to the thermodynamics of the van der Waals system' [arXiv:1205.3544v1], Quevedo and Ram\'{i}rez obtain the geodesics numerically by means of the equation $\ddot{E}^a+\Gamma^a_{~ bc} \dot{E}^b \dot{E}^c=0$ (where a dot denotes differentiation with respect to an arbitrary affine parameter); the Christoffel symbols are calculated from the thermodynamic metric on the space of equilibrium states. A different approach is taken in `A Lagrangian Description of Thermodynamics' [arXiv:1110.6152v1], where Vaz constructs a thermodynamic metric for several systems, including a Kerr black hole. In the case of the black hole, the equations for the temperature $T$ and angular velocity $\Omega$ take the role of equations of state, from which a metric is derived; the geodesic equations are determined from Hamilton's equations and it is pointed out that they can also be obtained by extremising the reparametrisation-invariant action.

The main aim of this work is to investigate the thermodynamic geodesics of an RN black hole in the space of equilibrium thermodynamic states. In a 2010 work by V\'{a}zquez, Quevedo and S\'{a}nchez, it was reported that there is no explicit time parameter in the GTD metric structures, and that the formalism did not as yet incorporate non-equilibrium thermodynamics \cite{Vazquez}. To our knowledge, this is still the situation at the present time. We therefore refrain from delving into finite-time thermodynamics and the associated dissipations. To this end, the black hole is modelled in such a way that its properties do not change significantly on a geometrical time scale, as further discussed in Sec. 3. 

The procedure we adopt is as follows: in Sec. 2 we derive a differential equation for the geodesics, and in Sec. 3 present a model of an RN black hole that evolves slowly via Hawking radiation and the Schwinger mechanism. We solve the geodesic equation numerically for this black hole in Sec. 4 and comment on the results, then conclude in Sec. 5. Throughout this paper, metrics are assigned the signature $(+,-,\dots,-)$ and, unless otherwise stated, the geometric unit system is adopted, with $G=c=k=k_\text{e}=1$ ($k_\text{e}$ is the Coulomb constant, equivalent to $1/4\pi\epsilon_0$). In these units, the reduced Planck's constant $\hbar$ becomes $2.6122\times 10^{-70}~\text{m}^2$ (to five significant figures and without the associated uncertainty)\footnote{The full value and its uncertainty, in SI units, are given in \cite{CODATA}}. Furthermore, the electric charge $Q$ of the black hole is assumed to be positive ($Q>0$).

All numerical analysis was carried out using $\text{Wolfram Mathematica}^{\text{\textregistered}} 10$. The figures were created using the \emph{LevelScheme} scientific figure preparation system \cite{LevelScheme}.

\section{A Differential Equation for the Geodesics}
\label{Differential Equations}

Thermodynamic systems characterised by second-order phase transitions, such as black holes, can be modelled as a contact manifold $T'$ with thermodynamic metric \cite{Bravetti2013}:   
\begin{equation}
G=(\text{d}\Phi-\delta_{ab}I^a \text{d}E^b)^2+(\delta_{ab}E^a I^b)(\eta_{cd}\text{d}E^c \text{d}I^d)
\end{equation}  
~~~~~~~~~\\
where $\delta_{ab}$ and $\eta_{ab}$ are the Euclidean and Minkowski metrics, respectively.

This in turn gives rise to the thermodynamic metric $g$ on $\varepsilon$ \cite{Bravetti2013}:
\begin{equation}
g=\left(E^a \frac{\partial\Phi}{\partial E^a}\right)\left(\eta_{b}^c\frac{\partial^2\Phi}{\partial E^c \partial E^d}\text{d}E^b \text{d}E^d\right)
\end{equation} 
The metric $g$ can easily be computed for a given thermodynamic system once the fundamental equation $\Phi=\Phi(E^a)$ is known \cite{Bravetti2013}. 

In the mass representation (i.e. with the mass acting as thermodynamic potential), the thermodynamic metric $g$ describing an RN black hole in a four-dimensional spacetime is given by:
\begin{equation}
[g_{ab}]=(SM_S+QM_Q)\left(
\begin{array}{ccc}
M_{SS} & 0 \\
0 & -M_{QQ}
\end{array}
\right)
\label{g}
\end{equation}
$M$ stands for the total mass of the black hole and $S$ its entropy, while $Q$ represents the electric charge. Subscripts denote \emph{partial} derivatives with respect to the corresponding coordinate. Eq. ($\ref{g}$) was obtained from \cite{Quevedo2008}, where it was used to describe an RN anti-de Sitter black hole, but it can easily be deduced that it is also valid in the absence of a cosmological constant. The metric signature was changed to $(+,~-)$. 

The Lagrangian $L$ takes the generic form $\sqrt{g_{ab}\dot{x}^a\dot{x}^b}$:
\begin{equation}
L=\sqrt{(SM_S+QM_Q)(M_{SS}\dot{S}^2-M_{QQ}\dot{Q}^2)}
\end{equation}
The dot stands for differentiation with respect to an arbitrary parameter $\zeta$ that is assumed to be affine. Substituting for $L$ in the Euler--Lagrange equations (where $x^1$ stands for $S$ and $x^2$ for $Q$):
\begin{equation}
\frac{\text{d}}{\text{d}\zeta}\left(\frac{\partial L}{\partial\dot{x}^a}\right)=\frac{\partial L}{\partial x^a};\qquad a=\{1,2\}
\end{equation}
then yields:
\begin{eqnarray}
\chi_S\dot{S}^2+2\chi_Q\dot{Q}\dot{S}+\xi_S\dot{Q}^2&=-2\chi\ddot{S}\label{geodesic1}\\
\chi_Q\dot{S}^2+2\xi_S\dot{Q}\dot{S}+\xi_Q\dot{Q}^2&=-2\xi\ddot{Q}\label{geodesic2}
\end{eqnarray}
with 
\begin{align}
\chi=&M_{SS}(SM_S+QM_Q);\notag\\ \xi=&M_{QQ}(SM_S+QM_Q)
\label{chi&xi}
\end{align}
Expressions for $M_S$, $M_Q$, $M_{SS}$ and $M_{QQ}$ can be obtained by first deriving an expression for $M$ from the Bekenstein--Hawking area--entropy relation, which reads \cite{Bekenstein, Bekenstein1973, Hawking1975}:
\begin{equation}
S=\frac{A}{4\hbar}
\label{Bekenstein-Hawking}
\end{equation}
The event--horizon area $A$ is computed as the surface area of a two-sphere with radius $r_+$, so that 
\begin{equation}
A=4\pi r_+^2
\label{A}
\end{equation}
where $r_+$ is the radius of the (outer) event horizon and is given by:
\begin{equation}
r_+=M+\sqrt{M^2-Q^2}
\label{radius}
\end{equation}
One can then simply write $A$ in terms of $M$ and $Q$ via ($\ref{A}$) and ($\ref{radius}$) and solve ($\ref{Bekenstein-Hawking}$) for $M$: 
\begin{equation}
M=\frac{\pi Q^2+\hbar S}{2\sqrt{\pi\hbar S}}
\label{M}
\end{equation}
This is equivalent to the Smarr mass formula \cite{Smarr} with the angular momentum set equal to zero, although \cite{Smarr} gives $M$ in terms of $A$ and $Q$ rather than $S$ and $Q$. 

The quantities $M_S$, $M_Q$, $M_{SS}$ and $M_{QQ}$ then follow easily from ($\ref{M}$), giving for $\chi$ and $\xi$ (Eq. ($\ref{chi&xi}$)):
\begin{equation}
\chi=\frac{9\pi^2 Q^4-\hbar^2 S^2}{32\pi\hbar S^3};~~~~~\xi=\frac{3\pi Q^2+\hbar S}{4\hbar S}
\label{chi&xi2}
\end{equation}

The geodesics in the space of equilibrium states can be determined by substituting for $\chi$, $\xi$ and their derivatives in ($\ref{geodesic1}$) and ($\ref{geodesic2}$) and solving the resulting differential equations. Nonetheless, a few comments are in order before we proceed. An RN black hole can be characterised by any two variables from the set $\{S, Q, M\}$. In this case, we have chosen the entropy $S$ and charge $Q$, with the third variable -- the mass $M$ -- uniquely determined by $S$ and $Q$ via ($\ref{M}$). However, given that the space of equilibrium states has coordinates $S$ and $Q$, any geodesic would have an equation of the form $f(S,Q)=0$. In other words, the geodesic equations introduce a dependence between $S$ and $Q$. Furthermore, as will be shown later, this dependence causes any changes in $M$ to be extremised. 

The starting point, therefore, is to write\footnote{Note that it is also possible to choose $Q$ as the dependent variable. This will be treated in greater detail in Sec. 4} the derivative $\dot{S}=\text{d}S/\text{d}\zeta$ as $\dot{S}=\text{d}S/\text{d}Q\times \text{d}Q/\text{d}\zeta=\text{d}S/\text{d}Q\times\dot{Q}$. Consequently, it becomes possible to combine ($\ref{geodesic1}$) and ($\ref{geodesic2}$) into one equation that reads (assuming $\dot{Q}\neq 0$):
\begin{align}
&\xi_S +\frac{\text{d}S}{\text{d}Q}\left[2\chi_Q-\frac{\chi}{\xi}\xi_Q\right]+\left(\frac{\text{d}S}{\text{d}Q}\right)^2 \left[\chi_S-\frac{2\chi}{\xi}\xi_S\right]\notag\\[0.5em]&-\left(\frac{\text{d}S}{\text{d}Q}\right)^3 \frac{\chi}{\xi}\chi_Q=-2\chi \frac{\text{d}^2 S}{\text{d}Q^2}
\label{main}
\end{align}
~~~~~~~~~\\
Substituting for $\chi$, $\xi$ (Eq. ($\ref{chi&xi2}$)) and the corresponding partial derivatives in ($\ref{main}$) yields:
\begin{align}
&-\frac{3\pi Q^2}{4\hbar S^2}+\frac{\hbar^2 S^2-3\pi Q^2(3\pi Q^2+2\hbar S)}{32\pi\hbar S^4}\left(\frac{\text{d}S}{\text{d}Q}\right)^2\nonumber\\[0.5em]&+\frac{3Q(9\pi Q^2+\hbar S)}{16 \hbar S^3}\frac{\text{d}S}{\text{d}Q}~=~\frac{\hbar^2 S^2-9\pi^2 Q^4}{16\pi \hbar S^3}\frac{\text{d}^2S}{\text{d}Q^2}\nonumber\\[0.5em]&+\frac{9Q^3(3\pi Q^2-\hbar S)}{64\hbar S^5}\left(\frac{\text{d}S}{\text{d}Q}\right)^3
\label{main2}
\end{align}
~~~~~~~~~~~~\\
The complexity of the differential equation thus obtained makes it exceedingly hard to solve analytically. Numerical techniques will instead be employed, but these require constraints which can only be determined by choosing an appropriate black hole model.

\section{Choosing a Black Hole Model}
\label{bhmodel}
In any astrophysically realistic case, charged black holes tend to get neutralised quickly. This happens because particles with an opposite charge are attracted to the black hole, neutralising some of its charge until this becomes too small to have a significant effect on the surrounding spacetime. Thus one begins by assuming that the black hole exists in isolation, surrounded by a perfect vacuum that is devoid of even cosmic background radiations \cite{Hiscock}. The assumption of complete isolation is perhaps not very plausible, but it becomes indispensable if an RN black hole with a geometrically interesting charge is to be investigated \cite{Hiscock}.  

Even if the black hole is not surrounded by any matter or radiation, it nonetheless discharges quickly due to the creation of electron--positron pairs in the electric field close to the horizon \cite{Hiscock}. This pair production would be rapid unless the mass of the black hole is very large ($> 10^5~\text{M}_{\odot}$) \cite{Gibbons}. Hence one makes the assumption that the black hole mass exceeds the said limit. This -- together with the isolation of the black hole -- allows the magnitude of the charge to be comparable to that of the black hole mass; the electric charge would then have a considerable effect on the geometry of spacetime \cite{Hiscock}. Very massive, charged black holes in isolation were considered by Hiscock and Weems in \cite{Hiscock} and in fact the above assumptions were made in accordance with their work. 

The only factors influencing the evolution of such a black hole would be Hawking radiation \cite{Hawking} and the Schwinger mechanism \cite{Schwinger}. We emphasise that this is only the case because the black hole is in isolation, and so can neither accrete matter from an external distribution nor absorb radiation. Although Hawking radiation actually refers to the emission of both massless and massive particles, stellar-mass black holes effectively emit only the former, their thermal energy being much less than the rest energy of massive particles \cite{Parentani}. This automatically excludes the production of charged particles, since these necessarily have mass. The same can be said of supermassive black holes. Charged-particle pairs are instead produced via the Schwinger mechanism. Strictly speaking, this mechanism should also be classified as a type of Hawking radiation \cite{Hiscock}. There is nonetheless a subtle difference between the two processes. In the case of Hawking radiation, the primary factor responsible for the separation of a virtual particle pair and the formation of real particles is the presence of a causal disconnection, while for the Schwinger mechanism it is the strong electric field surrounding the black hole. Both processes, however, lead to the slow evaporation of the black hole. For the purpose of this study, `slow' means that the mass and/or charge do not change significantly on a geometrical time scale $\tau$ ($\tau\simeq M$) \cite{Hiscock}. Thus the spacetime around the black hole can still be equipped with the usual RN metric, and its line element is given by
\begin{equation}
\text{d}s^2=f\text{d}t^2-f^{-1}\text{d}r^2\nonumber-r^2(\text{d}\theta^2+\sin^2{\theta}\phantom{|}\text{d}\phi^2)
\end{equation}
where
\begin{equation}
f\equiv\left(1-\frac{2M}{r}+\frac{Q^2}{r^2}\right)
\end{equation}
although $M$ and $Q$ should now be seen as slowly-varying functions of time \cite{Hiscock}.   
  
Hiscock and Weems make several other assumptions to construct their model. Three of these place a lower bound on the mass $M$ of the black hole, and they can be summarised as the requirement that $M>>Q_0$, where $Q_0=e\hbar/\pi m_{\text{e}}^2\approx \num{1.7e5}~\text{M}_\odot$ ($e$ being the elementary charge and $m_\text{e}$ the electron mass). This justifies the use of flat-space quantum electrodynamics and the truncation of the Schwinger formula (presented below) to the first term. Furthermore, since particles created with a charge of the same sign as the black hole's are acted upon by a very large radial repulsive force, and those with an opposite charge are absorbed by the black hole, scattering can effectively be neglected. Under these approximations, the rate $\Gamma$ per unit four-volume at which particles of mass $m$ are produced in pairs can be represented by the Schwinger formula (\cite{Schwinger} as cited by \cite{Hiscock}):
\begin{align}
&\Gamma =\notag\\ &\frac{e^2}{4\pi^3\hbar^2}\frac{Q^2}{r^4}\exp{\left(-\frac{\pi m^2 r^2}{\hbar e Q}\right)\left(1+O\left[\frac{e^3 Q}{m^2 r^2}\right]+\dots\right)}
\label{Schwinger}
\end{align}
and it is possible to find $\text{d}Q/\text{d}t$ by integrating $\Gamma$ over the three-volume outside $r_+$. First, however, the temporal dimension of the four-volume should be re-expressed in terms of the coordinate time, rather than the proper time \cite{Hiscock}. Upon integration, one obtains an expression for $\text{d}Q/\text{d}t$ with a term in $\text{erfc}[r_+/\sqrt{Q Q_0}]$, where $\text{erfc}(x)$ denotes the complementary error function of $x$. If the condition that $M>>Q_0$ is satisfied, this function can be approximated by its asymptotic series \cite{Hiscock}. The rate of charge loss from the black hole is then given by \cite{Hiscock}:
\begin{equation}
\frac{\text{d}Q}{\text{d}t}=-\frac{1}{2\pi^3}~\frac{e^4}{\hbar m_{\text{e}}^2}~\frac{Q^3}{r_+^3}\exp{\left(-\frac{r_+^2}{Q_0Q}\right)}
\label{dQdt}
\end{equation}

Since $\text{d}Q/\text{d}t$ has an exponential dependence on the square of the mass of the created particles in the denominator (via $Q_0$), the contributions of muons and heavier particles were ignored \cite{Hiscock}. 

The evaporation of the charged black hole results in mass being lost at a total rate: 
\begin{equation}
\frac{\text{d}M}{\text{d}t}=-\sigma\varepsilon T^4 A + \frac{Q}{r_+}\frac{\text{d}Q}{\text{d}t}
\label{dMdt}
\end{equation}
where $\varepsilon$ is the emissivity, $T$ the Hawking temperature and $A$ the event--horizon area. The Stefan--Boltzmann constant $\sigma$ is given by:
\begin{equation}
\sigma=\frac{\pi^2}{60\hbar^3}
\label{SB}
\end{equation}
Since a black hole is modelled as a black body, $\varepsilon$ is set equal to one.

The first term on the right-hand side of ($\ref{dMdt}$) is simply the radiated power as stipulated by the Stefan--Boltzmann law, and accounts for the energy lost as a result of the emission of massless particles. It is a slightly modified version of the expression in \cite{Hiscock}. Hiscock and Weems model the rate of mass loss due to thermal emission in terms of the total cross section of the black hole, while taking into account -- via a parameter $\alpha$ -- the number of neutrino species produced and the thermally-averaged cross sections for neutrinos, photons and gravitons. The use of their expression in ($\ref{dMdt}$) instead of $-\sigma\varepsilon T^4 A$ simply amounts to replacing $r_+^2$ (which determines $A$ according to ($\ref{A}$)) with the product of $\alpha$ and $b_\text{c}^2$, the squared critical value of the apparent impact parameter for photons, and did not yield any significant differences in our results. In fact, if $\alpha$ is fixed at\footnote{Two values are given in \cite{Hiscock} -- one for the emission of three massless neutrino species, and another (\num{0.26792}) valid when no massless neutrinos are produced, where by massless is meant a mass less than about \SI{e-10}{eV} (the black hole would be too cold to emit anything heavier) \cite{Hiscock}. Since the upperbounds available nowadays for the mass of neutrino flavours \cite{ParticleDataGroup} are much greater than this value, we use $\alpha=\num{0.26792}$ when comparing our work with \cite{Hiscock}} \num{0.26792}, the ratio of $\alpha b_\text{c}^2$ to $r_+^2$ for the values set down in ($\ref{ic}$) is close to unity, meaning that both approaches yield the same order of magnitude for the thermal power. Hence we opt for the Stefan--Boltzmann law, this being considerably simpler and, in fact, quite popular in the literature (see, for instance, \cite{Hobson, Thornton, Walecka}). The second term on the right-hand side of ($\ref{dMdt}$) arises due to pair production via the Schwinger mechanism.   
 
It now becomes necessary to consider the first law of black hole thermodynamics as applied to an RN black hole:
\begin{equation}
\text{d}M=T\text{d}S+\phi\phantom{|}\text{d}Q
\label{firstlaw}
\end{equation}
Equivalently:
\begin{equation}
\frac{\text{d}M}{\text{d}t}=T\frac{\text{d}S}{\text{d}t}+\phi\frac{\text{d}Q}{\text{d}t}
\label{dMdt2}
\end{equation}
~~~~~~~~~\\
and since the electrostatic potential $\phi$ is given by $Q/r_+$, it can easily be deduced, by comparing ($\ref{dMdt2}$) with ($\ref{dMdt}$), that
\begin{equation}
T\frac{\text{d}S}{\text{d}t}=-\sigma T^4 A
\label{comparison}
\end{equation}
The Hawking temperature $T$ is related to the surface gravity $\kappa$ via the equation $T=(\hbar\kappa)/2\pi$ \cite{Hawking, Hawking1975}, but it can also be obtained by considering Eq. ($\ref{firstlaw}$). Since $M$ is a state function, d$M$ must be an exact differential, from which it follows that $\partial M/\partial S=T$ (and similarly, $\partial M/\partial Q=\phi$). 
Eq. ($\ref{M}$) then gives for $T$:
\begin{equation}
T=\frac{-\pi Q^2+\hbar S}{4\sqrt{\pi \hbar S^3}}
\label{T}
\end{equation}
The radius of the event horizon can be written as a function of $S$ by substituting for $M$ (Eq. ($\ref{M}$)) in ($\ref{radius}$):
\begin{equation}
r_+=\sqrt{\hbar S/\pi}
\label{rnew}
\end{equation}
so that Eq. ($\ref{A}$) becomes $A=4\hbar S$. This can be inserted, together with equations ($\ref{SB}$) and ($\ref{T}$), into ($\ref{comparison}$) to obtain an expression for the rate of change of entropy:
\begin{equation}
\frac{\text{d}S}{\text{d}t}=-\frac{\sqrt{\pi}(\hbar S-\pi Q^2)^3}{960(\hbar S)^{7/2}}
\label{dSdt}
\end{equation}
while substituting ($\ref{rnew}$) for $r_+$ in ($\ref{dQdt}$) and further simplication results in an expression for the rate of charge loss in terms of $S$ and $Q$:
\begin{equation}
\frac{\text{d}Q}{\text{d}t}=-\frac{e^4 \exp{\left(-\frac{m_{\text{e}}^2 S}{eQ}\right)}Q^3}{2m_{\text{e}}^2 \sqrt{\pi^3 \hbar^5 S^3}}
\label{dQdt2}
\end{equation}

\section{Choosing Appropriate Constraints and Solving}
\label{constraints}
From ($\ref{dSdt}$) and ($\ref{dQdt2}$) it follows that:
\begin{align}
\frac{\text{d}S}{\text{d}Q}&=\frac{\text{d}S}{\text{d}t}\frac{\text{d}t}{\text{d}Q}\notag\\&=-\frac{\pi^2\hbar^{5/2}m_{\text{e}}^2 S^{3/2}(\pi Q^2-\hbar S)^3 \exp{\left(\frac{m_{\text{e}}^2S}{eQ}\right)}}{480e^4Q^3(\hbar S)^{7/2}}
\label{dSdQ}
\end{align}

The elementary charge $e$ and electron mass $m_{\text{e}}$ have values \SI{1.3807e-36}{m} and \SI{6.7646e-58}{m}, respectively (each stated to five significant figures and without the associated uncertainty)\footnote{The full values and uncertainties, in SI units, are given in \cite{CODATA}}. 

The necessary constraints were obtained by fixing the mass $M$ at $M_*=\SI{3e9}{m}$ (\num{2e6}~$\text{M}_\odot$), with $Q_*$ set equal to \SI{8E8}{m} (\SI{9e25}{C}). The latter satisfies the requirement that the charge be comparable in magnitude to the mass of the black hole. For the given value of $M_*$, the maximum charge the black hole can have is also \SI{3e9}{m}, in which case the black hole would be extremal. This is equivalent to \SI{3e26}{C}.

The entropy $S$ can be calculated from ($\ref{Bekenstein-Hawking}$) after substituting for $A$ (Eq. ($\ref{A}$)) and $r_+$ (Eq. ($\ref{radius}$)). When $M_*=\SI{3e9}{m}$ and $Q_*=\SI{8e8}{m}$, $S_*$ evaluates to \num{4.1742e89}, while Eq. ($\ref{dSdQ}$) yields a value for $(\text{d}S/\text{d}Q)_*$ of \SI{1.7311e26}{m^{-1}}. 

\begin{figure}
\includegraphics[width=\columnwidth]{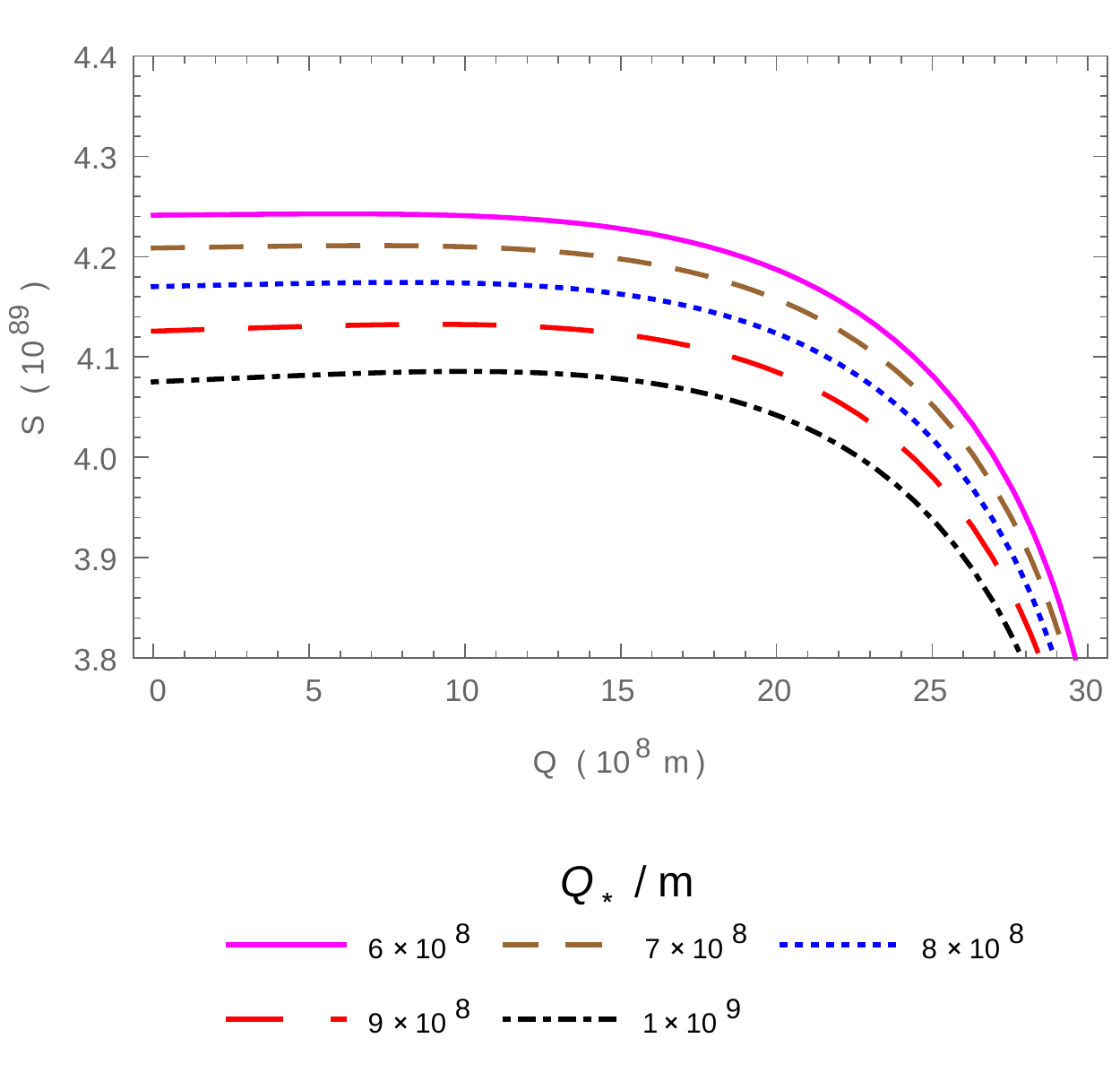}
\caption{\label{rnbh1}Thermodynamic geodesics obtained by solving Eq. $(\ref{main2})$ numerically for 5 different sets of constraints, each calculated at $M_*=\SI{3e9}{m}~(\num{2e6}~\text{M}_\odot)$}
\end{figure}

Thus the constraints used to solve ($\ref{main2}$) numerically are as follows:
\begin{eqnarray}
Q_*=\SI{8e8}{m}&;~~~ S_*=\num{4.1742e89};\nonumber\\ (\text{d}S/\text{d}Q)_*&=~\SI{1.7311e26}{m^{-1}}
\label{ic}
\end{eqnarray}
Similarly, four other sets of constraints were obtained, each corresponding to the same $M_*$ but a different value of $Q_*$. The resulting solutions, together with the one for $(\ref{ic})$, are illustrated in Fig. $\ref{rnbh1}$. 
\begin{figure*}
\captionsetup[subfigure]{oneside,margin={1.1cm,0cm}}
\centering
\subfloat[\label{numericala}]{\includegraphics[width=70mm]{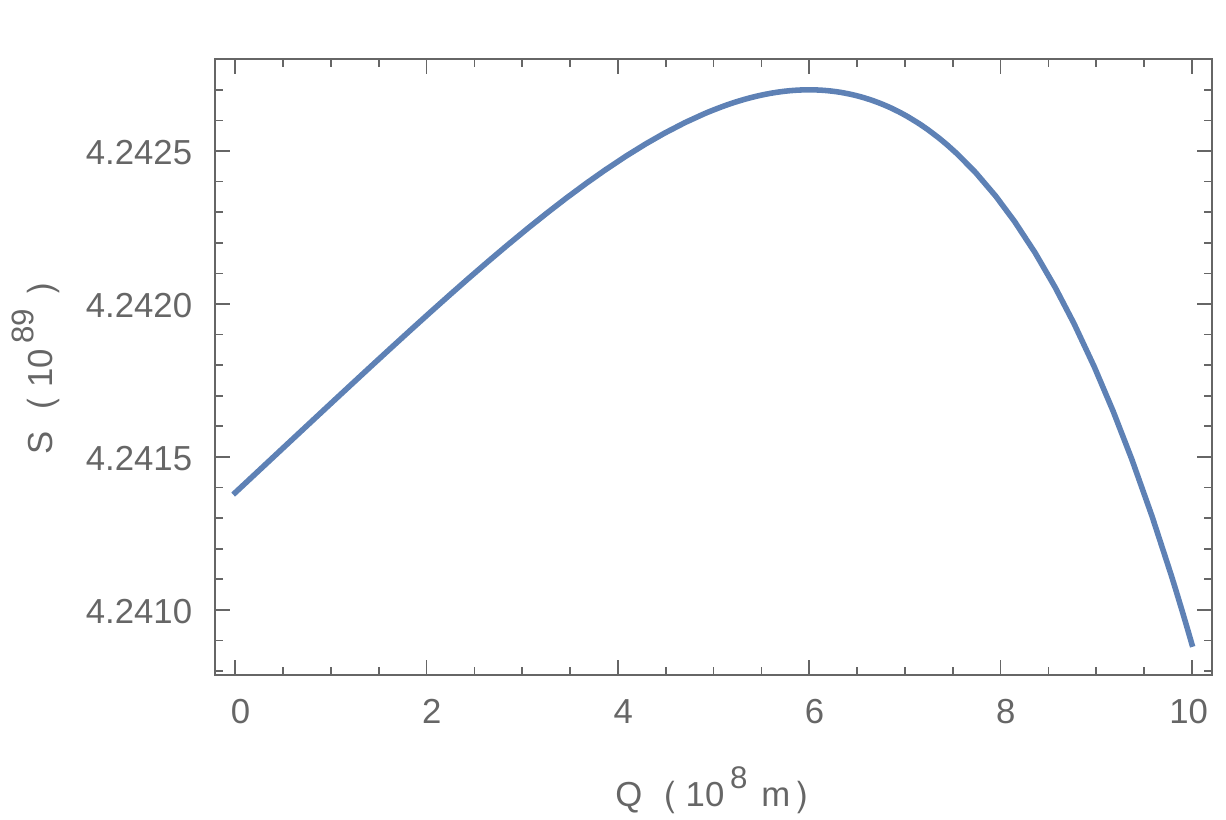}}\hspace{0.5cm} 
\subfloat[\label{numericalb}]{\includegraphics[width=70mm]{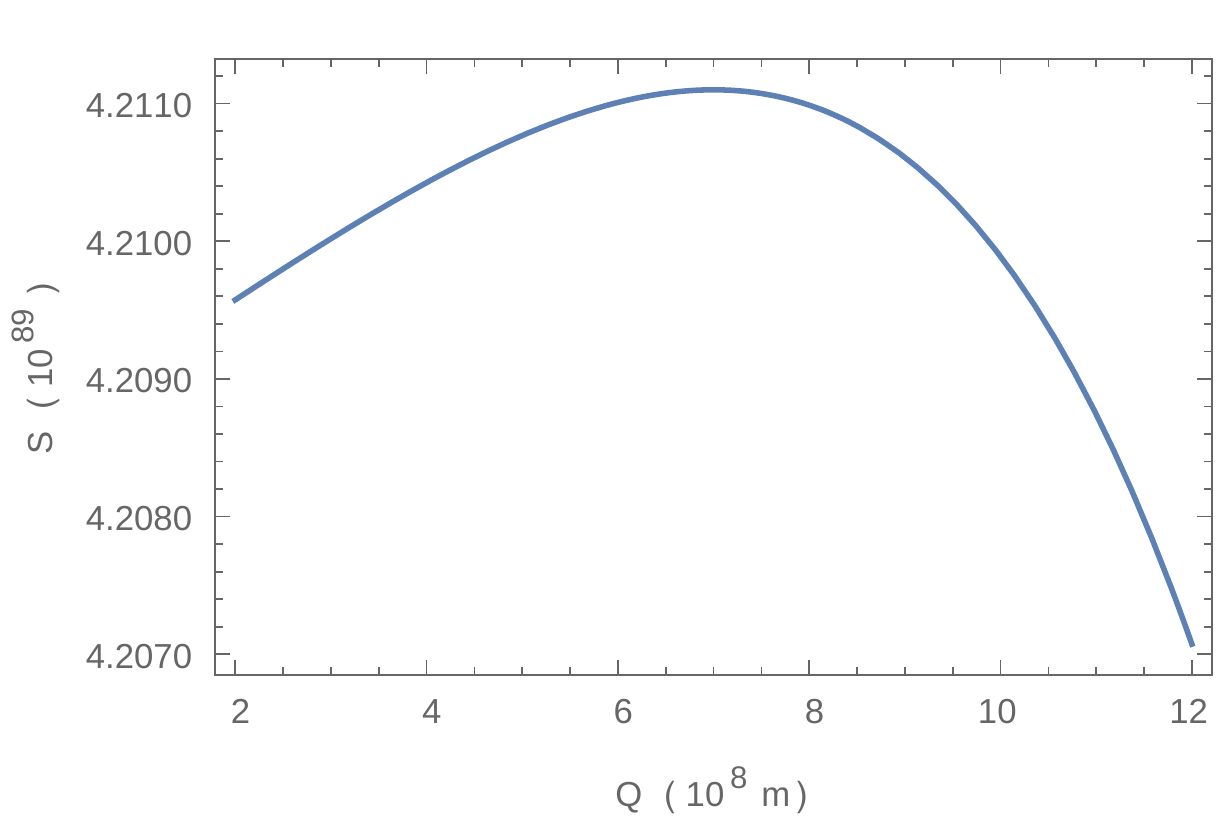}}\\
\subfloat[\label{numericalc}]{\includegraphics[width=70mm]{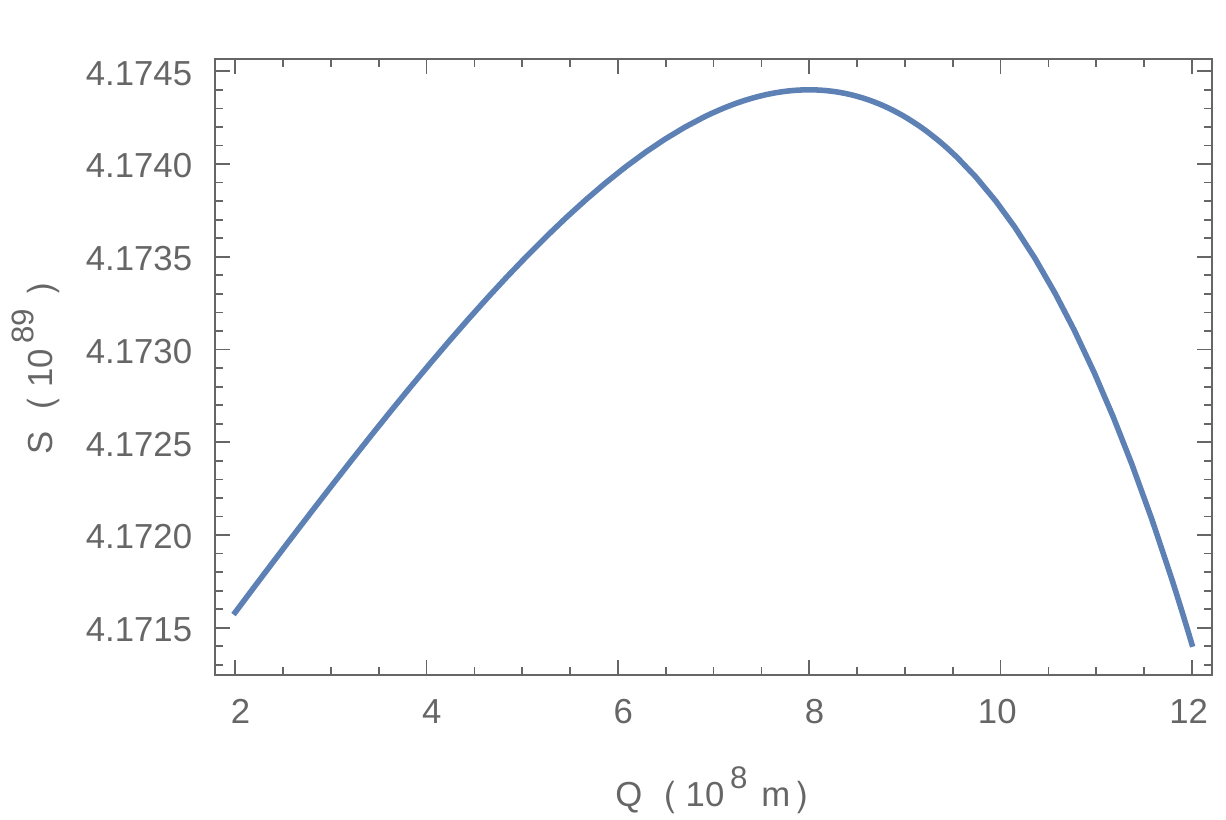}}\hspace{0.5cm}
\subfloat[\label{numericald}]{\includegraphics[width=70mm]{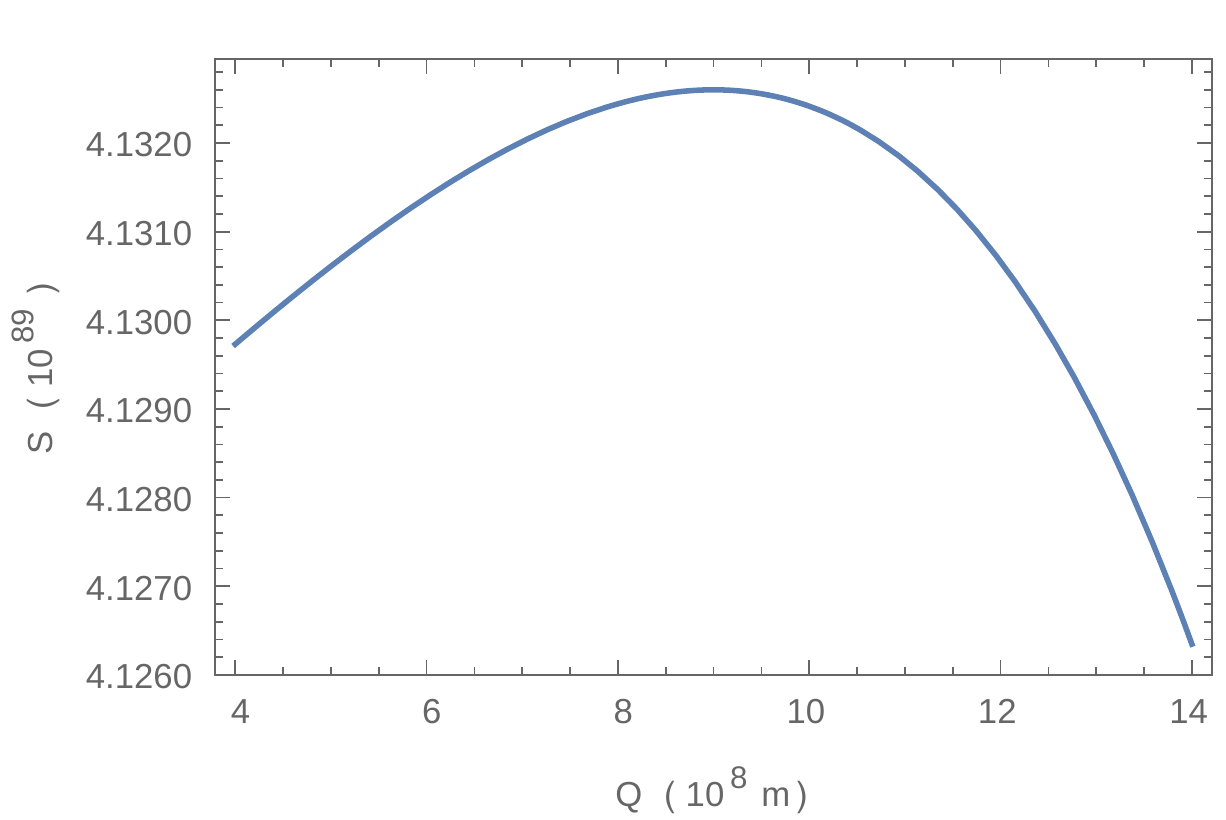}}\\
\subfloat[\label{numericale}]{\includegraphics[width=70mm]{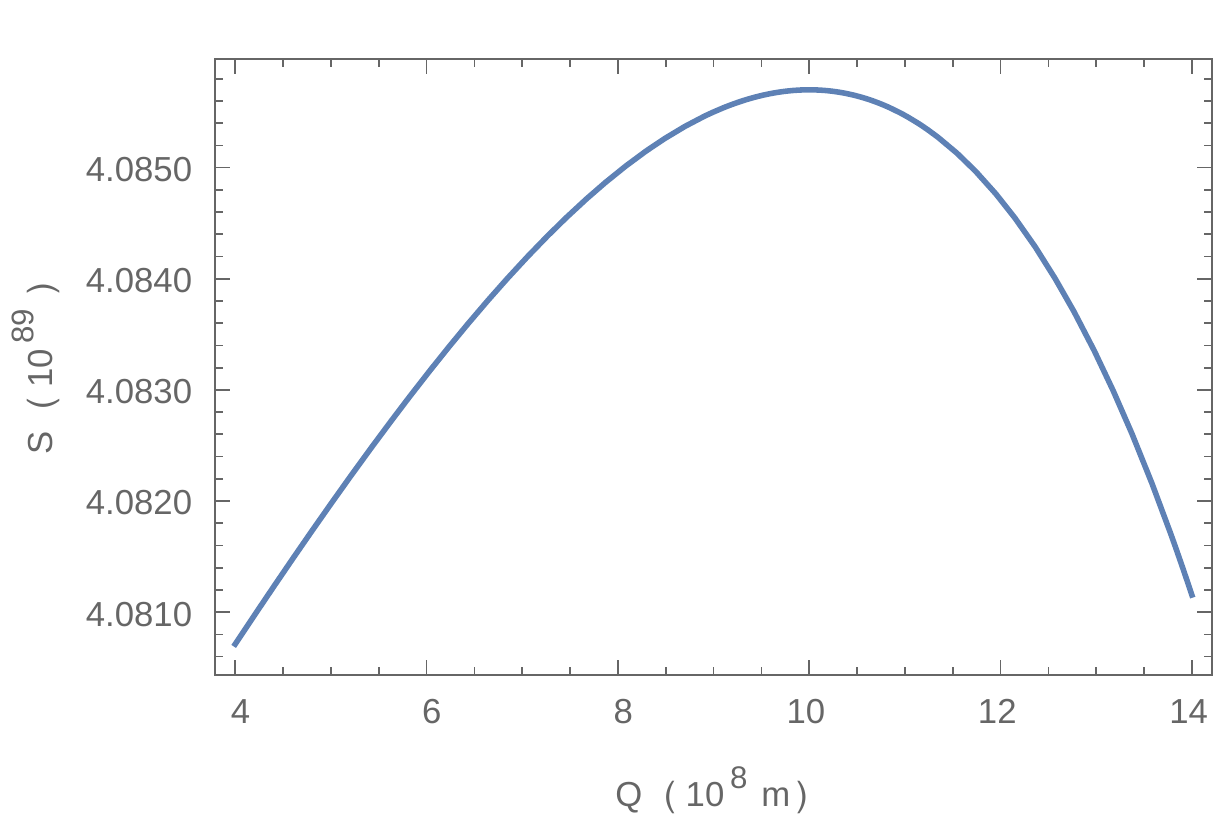}}\vspace{1.5cm}
\caption{\label{maxima}Thermodynamic geodesics over a restricted domain. $M_*$ is set to \SI{3e9}{m}, with $Q_*$ being: \textbf{(a)} \num{6e8} ~~\textbf{(b)} \num{7e8} \textbf{(c)} \num{8e8} \textbf{(d)} \num{9e8} and \textbf{(e)} \num{1e9} metres} 
\end{figure*} 
A closer inspection (Fig. $\ref{maxima}$) reveals that each trajectory is characterised by a maximum that occurs very close to the state specified by the constraints, indicating that -- to a good approximation -- a black hole with $M=M_*$ and $Q=Q_*$ evolves along the corresponding geodesic by losing entropy via Hawking radiation, both if discharging and also in the purely theoretical scenario of an increase in $Q$. This can be traced back to the fact that the Schwinger mechanism does not alter the entropy of the black hole, as can be demonstrated by a simple calculation. One should first note that for each pair of particles produced, the charge of the black hole decreases by $e$ and the mass\footnote{The rest energy of the positron (the particle with the same-sign charge as the black hole) is not taken into account; the large charge-to-mass ratio of this particle implies that the energy $eQ/r_+$ it gains when repelled to infinity is much greater \cite{Hiscock}} by $eQ/r_+$. Using the equation:
\begin{equation}
\Updelta r_+=\Updelta M\left(\frac{M+\sqrt{M^2-Q^2}}{\sqrt{M^2-Q^2}}\right)-\Updelta Q\frac{Q}{\sqrt{M^2-Q^2}}
\end{equation} 
in conjunction with the relations $\Updelta M=-eQ/r_+$ and $\Updelta Q=-e$ yields the result $\Updelta r_+=0$. Hence, the area of the event horizon (Eq. ($\ref{A}$)) remains constant and, as follows from Eq. ($\ref{Bekenstein-Hawking}$), so does the entropy.  
\begin{figure*}
\captionsetup[subfigure]{oneside,margin={0.85cm,0cm}}
\subfloat[]{\includegraphics[width=70mm]{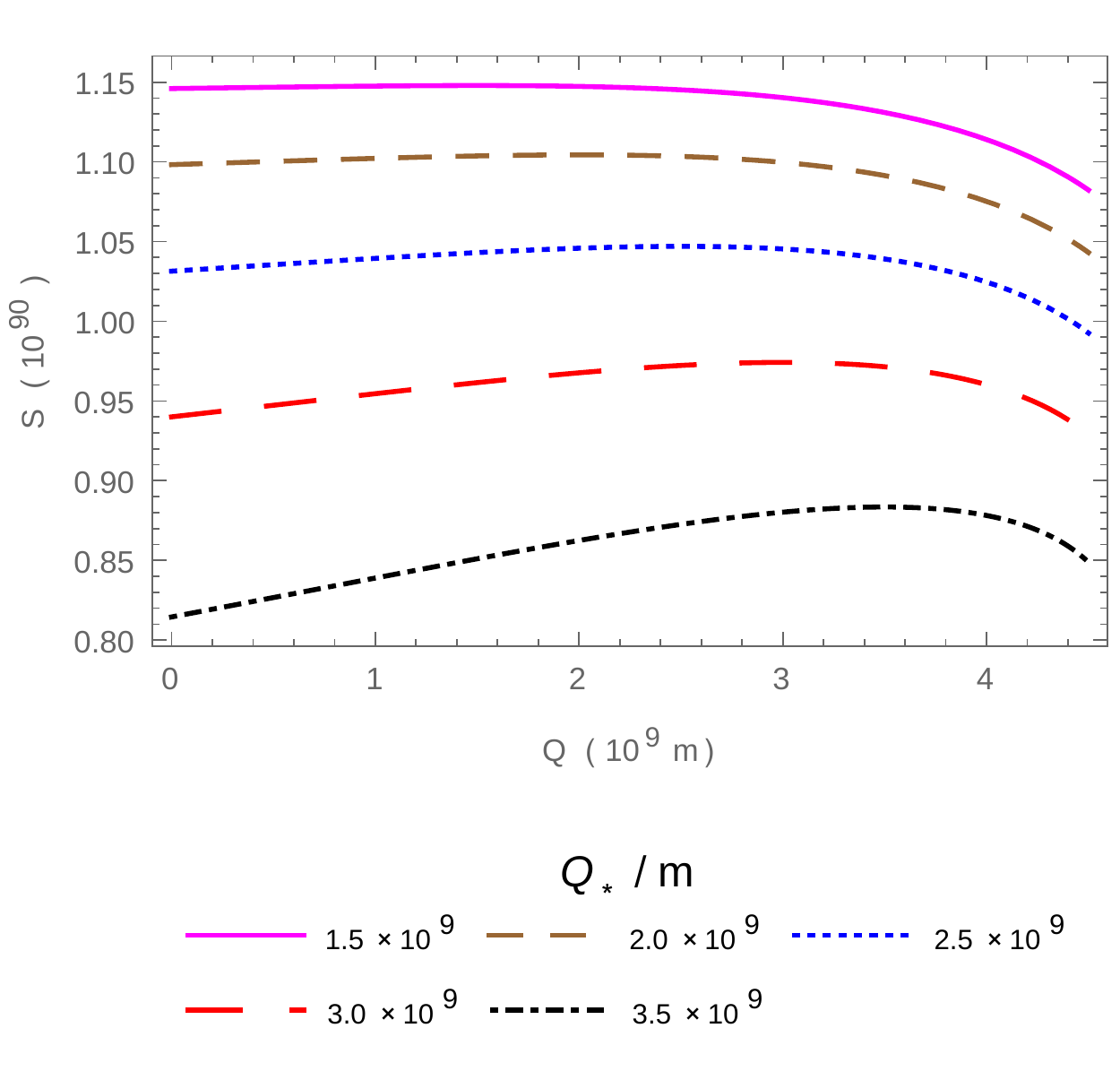}}\hspace{0.9cm} 
\subfloat[]{\includegraphics[width=70mm]{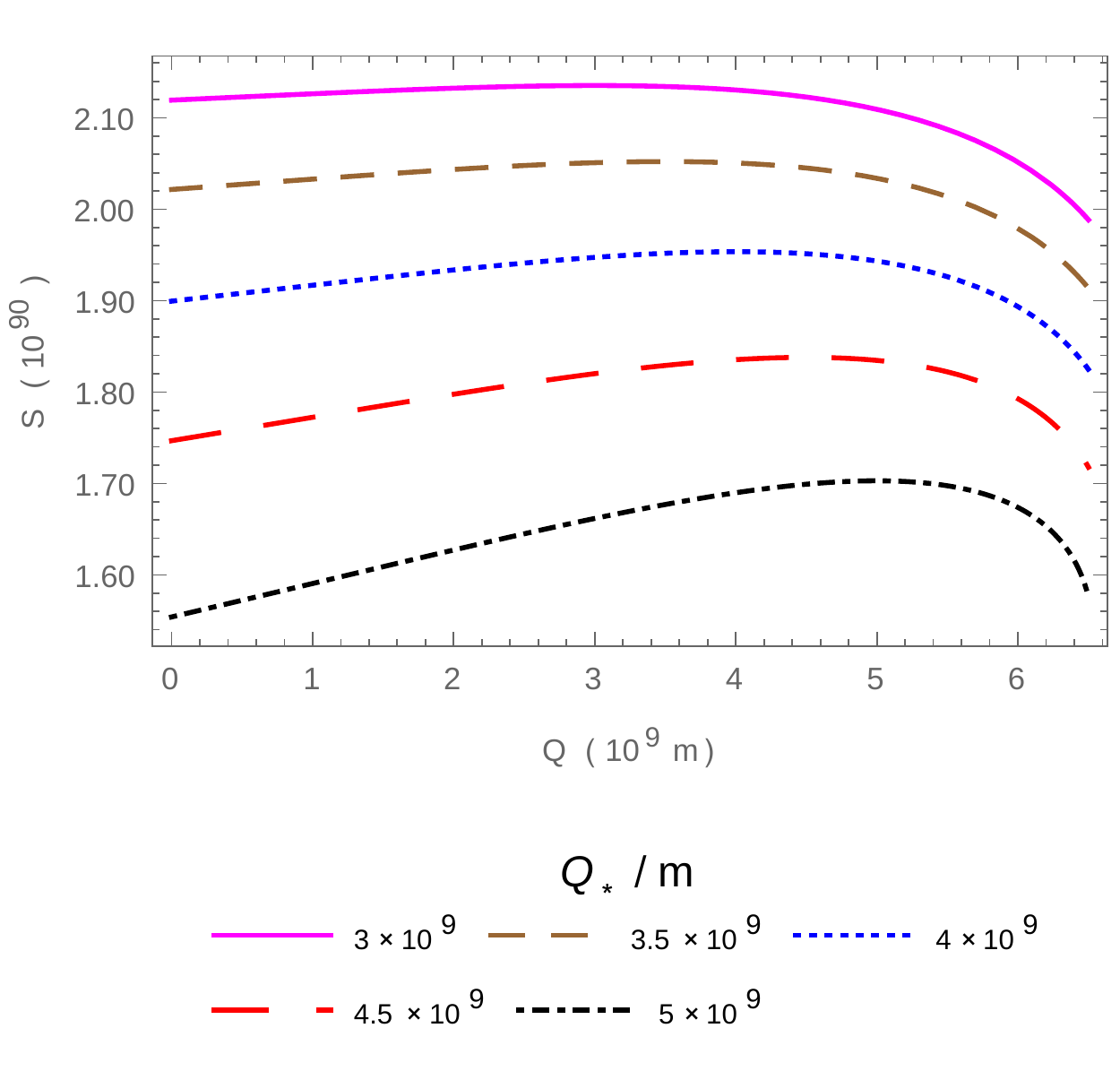}}
\caption{Thermodynamic geodesics obtained by solving Eq. $(\ref{main2})$ numerically for 5 different sets of constraints, each calculated at \textbf{(a)} $M_*=\SI{5e9}{m}$ (\num{3.4e6} $\text{M}_\odot$) and \textbf{(b)} $M_*=\SI{7e9}{m}$ (\num{4.7e6} $\text{M}_\odot$)} 
\label{geodesics} 
\end{figure*} 

One should also note that the closer $Q_*$ is to the extremal value, the smaller the value of $S_*$ and consequently, the lower the corresponding curve. Since the five sets of constraints were obtained for the same value of $M_*$, this mirrors the fact that at a fixed value of $M$ ($=M_*$), the entropy ($S_*$) is least when the black hole is closest to extremality. Furthermore, each curve bends downwards as $Q$ approaches its extremal value, so that the entropy decreases and is least for maximum $Q$. In the limit of extremality, however, Eq. ($\ref{main2}$) becomes stiff. This is in line with the third law of black hole thermodynamics, which forbids a non-extremal black hole from evolving into an extremal one. Strictly speaking, an extremal black hole is defined as having $M$ equal to $Q$ (in the geometric unit system), and given that $M$ evolves from its constraining value $M_*$ along the geodesic, the reader should be aware that the word `extremality' is used rather loosely in this situation, since here an extremal black hole is considered to be one with $Q=M_*$. 

Fig. $\ref{rnbh1}$ also shows that at small values of $Q$, the geodesics can be well-approximated by straight lines. This indicates that the space of equilibrium states becomes less curved, and the thermodynamic activity of the black hole decreases. The possibility of using curvature as a probe of the thermodynamic interactions of a system makes GTD very advantageous, as outlined in the Introduction. In fact, the equilibrium states of an ideal gas, whose constituent particles do not interact, give rise to a flat space. Consequently, with the proper choice of coordinates, the resulting geodesics take the form of straight lines \cite{Quevedo2015}. 

It should nonetheless be noted that, as shown by Hiscock and Weems, it is possible for sufficiently large RN black holes to initially evolve closer to extremality by shedding their mass while keeping their charge approximately constant. This is unlike what happens in the case of a rotating black hole, which always evolves towards the Schwarzschild limit \cite{Hiscock}.

The analysis was repeated for two other constraining values of $M$, and the results are presented in Fig. $\ref{geodesics}$. In all three cases, the constraints were not chosen as randomly as it might seem, because certain values for $M_*$, despite being significantly greater than $Q_0$ ($\approx \num{1.7e5}~\text{M}_\odot$), yield very high rates of change for values $Q_*$ whose order of magnitude is comparable to that of $M_*$, implying a black hole that would discharge much too quickly to be of relevance to this work. In most of these situations, $(\text{d}S/\text{d}t)_*$ also turns out to be unacceptably large.

The fact that $S$ always decreases for a black hole with $Q=Q_*$ and $S=S_*$, regardless of what $Q$ does, might initially seem at odds with the assumption that $S$ is a function of $Q$. However, although the Schwinger mechanism does not change the area $A$ of the event horizon, it \emph{does} affect $\text{d}A/\text{d}t$. Indeed, the loss of charge makes $A$ decrease faster, and thus effectively increases the rate at which entropy is lost. This becomes apparent if one takes the time derivative of the relation $A=4\hbar S$ (the Bekenstein--Hawking area--entropy relation, Eq. ($\ref{Bekenstein-Hawking}$)) and substitutes for $\text{d}S/\text{d}t$ (Eq. ($\ref{dSdt}$)), getting:
\begin{equation}
\frac{\text{d}A}{\text{d}t}=-4 \hbar \frac{\sqrt{\pi}(\hbar S-\pi Q^2)^3}{960(\hbar S)^{7/2}}
\end{equation}
Given that the Schwinger mechanism leaves the entropy intact but decreases the charge, the overall result is an increase in $|\text{d}A/\text{d}t|$ and hence in the rate of entropy loss. In other words, changes in $Q$ have a direct bearing on the amount by which $S$ decreases in a given time. It is in this sense that the geodesics in the space of thermodynamic equilibrium states can be expressed as a function $f(S,Q)=0$. Furthermore, one might just as well have chosen $Q$ as the dependent variable, because any function $Q(S)$ has a corresponding inverse $S(Q)$, provided it is one-to-one. Should a solution $Q(S)$ to the geodesic equation (this having been expressed in terms of $Q'(S)$ and $Q''(S)$) be many-to-one, its inverse -- or a portion of it -- would still turn up as a solution of Eq. ($\ref{main2}$), but the gradient of the resulting curve would be singular at one or more points, and the solver would detect a stiffness problem there. Fortunately, the solutions presented in this work are not stiff over the given domain, and so the possibility that the technique employed might be generating portions of one-to-many solutions is eliminated.  
    
We now turn our attention to the physical significance of the result, in the sense that we shall try to understand how our analysis fits in with the evolution of a `real' black hole, whose thermodynamics would be governed at all times by equations ($\ref{dQdt}$) and ($\ref{dMdt}$). The starting point is the metric on the space of thermodynamic equilibrium states (Eq. ($\ref{g}$)). We write it again below for ease of reference: 
\begin{equation}
[g_{ab}]=(SM_S+QM_Q)\left(
\begin{array}{ccc}
M_{SS} & 0 \\
0 & -M_{QQ}
\end{array}
\right)
\label{g2}
\end{equation}
The two-by-two matrix on the right-hand side is reminiscent of the metric proposed by Weinhold \cite{Weinhold}, which can be expressed as $g_\text{W}=\partial^2 U/\partial E^\alpha\partial E^\beta$ ($E^\alpha$ and $E^\beta$ represent extensive variables). Weinhold's metric can provide a measure of distance in several ways. As mentioned in the Introduction, Weinhold himself worked in the tangent space defined at a general point of the equilibrium manifold \cite{Andresen, Gilmore}. Let us suppose that our $n$-dimensional manifold -- described by the equation of state $U=U(E^\alpha)=U(E^1,\dots,E^n)$ -- is embedded in an $(n+1)$-dimensional space $\mathbb{R}^{n+1}$. Its coordinates in this space would be $(E^\alpha;U(E^\alpha))$. If we then consider the tangent space at an equilibrium point $(E_0^\alpha; U(E_0^\alpha))$ and adopt Weinhold's choice of metric, the quantity $\Updelta s^2=g_\text{W}\Updelta E^\alpha \Updelta E^\beta$ would denote the square of the distance between $(E_0^\alpha; U(E_0^\alpha))$ and the neighbouring point \emph{in the tangent space} with coordinates\footnote{Here, $\lambda_\alpha$ stands for the intensive thermodynamic variables of the system. At equilibrium, $\lambda_\alpha=\frac{\partial U}{\partial E^\alpha}\Bigr|_{\substack{E_0^\alpha}}$ \cite{Gilmore}} $(E_0^\alpha+\Updelta E^\alpha; U(E_0^\alpha)+\lambda_\alpha \Updelta E^\alpha)$ \cite{Gilmore}. To lowest order, $g_\text{W}\Updelta E^\alpha \Updelta E^\beta$ is also equivalent to twice the distance between the displaced state $(E_0^\alpha+\Updelta E^\alpha; U(E_0^\alpha+\Updelta E^\alpha))$ \emph{on the manifold} and the tangent space at the equilibrium point $(E_0^\alpha; U(E_0^\alpha))$ \cite{Salamon1980}. In the latter case, the distance can be identified with the availability of the displaced system \cite{Salamon1980}.

Additionally, it is possible to use Weinhold's metric as a means of introducing a notion of distance in the equilibrium manifold itself, although the Gibbsian picture of the space of equilibrium states as a convex hypersurface would then have to be abandoned \cite{Andresen}. The authors of \cite{Nulton} concluded that the thermodynamic length thus computed controls the dissipation in finite-time processes. This approach is more relevant to us, since we are specifically interested in thermodynamic distances in the space of equilibrium states.  

Both of the above-mentioned applications show that Weinhold's metric makes it possible to obtain bounds on the change in energy associated with a thermodynamic process. The main difference between Weinhold's metric and the two-by-two matrix in Eq. ($\ref{g2}$) is the introduction in the latter of the metric signature $(+,-)$. Just as in normal spacetime, the signature distinguishes between the temporal dimension and the spatial ones, its role in the line element constructed from ($\ref{g2}$) is to make the distinction between the term in $\text{d}Q^2$, which represents fluctuations in the electrostatic energy of the black hole, and that in $\text{d}S^2$. The latter is associated with changes in the irreducible mass of the black hole, so called because it cannot be extracted by any classical process. 

The two-by-two matrix in Eq. ($\ref{g2}$) is scaled by the sum $SM_S+Q M_Q$, equivalent to $ST+Q\phi$. If we compare this sum to the first law of black hole thermodynamics (Eq. ($\ref{firstlaw}$)), it can easily be deduced that it amounts to the total energy (mass) expended by an `infinite reservoir' -- whose temperature and electrostatic potential remain constant at all times -- to produce a black hole having entropy $S$ and charge $Q$.

These considerations allow us to conclude that extremising the thermodynamic length computed from ($\ref{g2}$) must yield some important information about the way the black hole's mass changes as it evolves. To further clarify the nature of this information, we extend our analysis to the geodesic equation itself. It turns out that the expression for the entropy of an extremal black hole -- i.e. $S=\pi Q^2/\hbar$ -- is an exact solution to Eq. ($\ref{main2}$). Let us consider a point $P$ on this curve. A black hole at $P$, being extremal, cannot increase its charge unless it first gains mass. The smallest increase in mass occurs if the black hole remains extremal as it evolves -- i.e. if it `moves' to a state $P+\delta P$ that also lies on the curve $S=\pi Q^2/\hbar$. On the other hand, should the black hole at $P$ lose charge, it could only keep to the curve if the decrease in $M$ is the largest possible. In other words, the curve $S=\pi Q^2/\hbar$ is a solution to the geodesic equation which extremises the change in mass accompanying a variation in the electric charge of a hypothetical black hole, maximising $\Updelta M$ if the black hole is discharging and minimising it if $Q$ increases. It can thus be inferred, irrespective of whether the solution $S=\pi Q^2/\hbar$ has physical meaning\footnote{Several authors are of the opinion that a black hole which is exactly extremal has zero entropy; see, for instance, \cite{Hawking1995, Hod, Teitelboim, Carroll, Das}}, that \emph{any} solution to the geodesic equation extremises the change in $M$ for a given increase or decrease in $Q$. 

\begin{figure*}
\captionsetup[subfigure]{oneside,margin={0.82cm,0cm}}
\centering
\subfloat[\label{}]{\includegraphics[width=70mm]{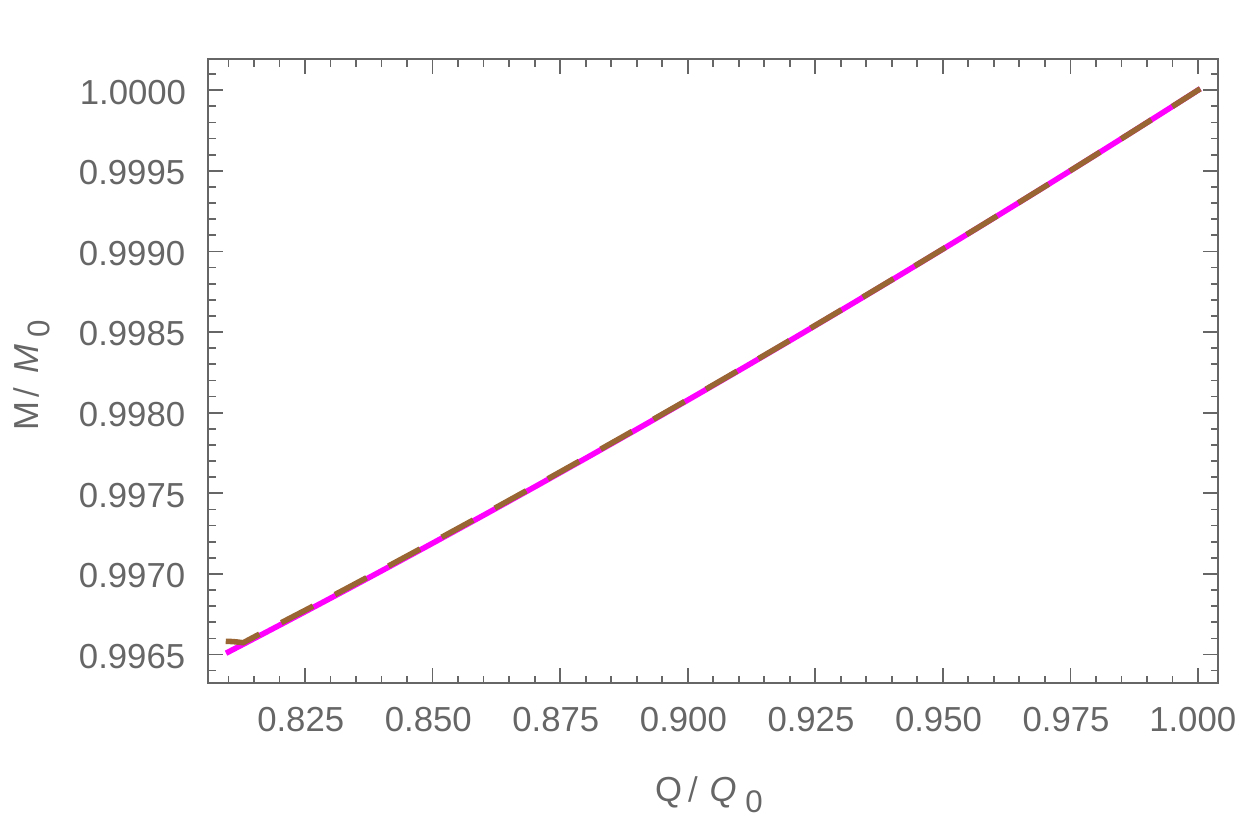}}\hspace{0.5cm} 
\subfloat[\label{}]{\includegraphics[width=70mm]{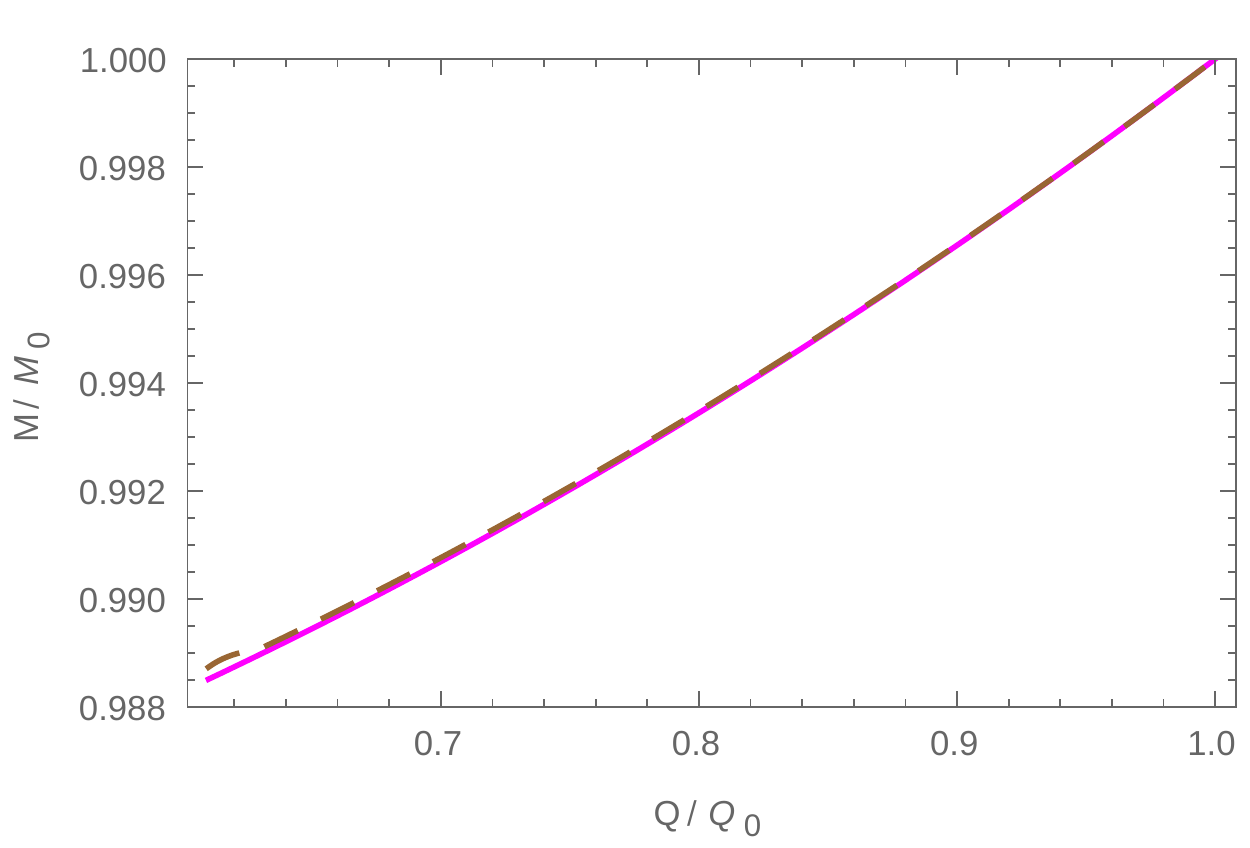}}\\
\subfloat[\label{}]{\includegraphics[width=70mm]{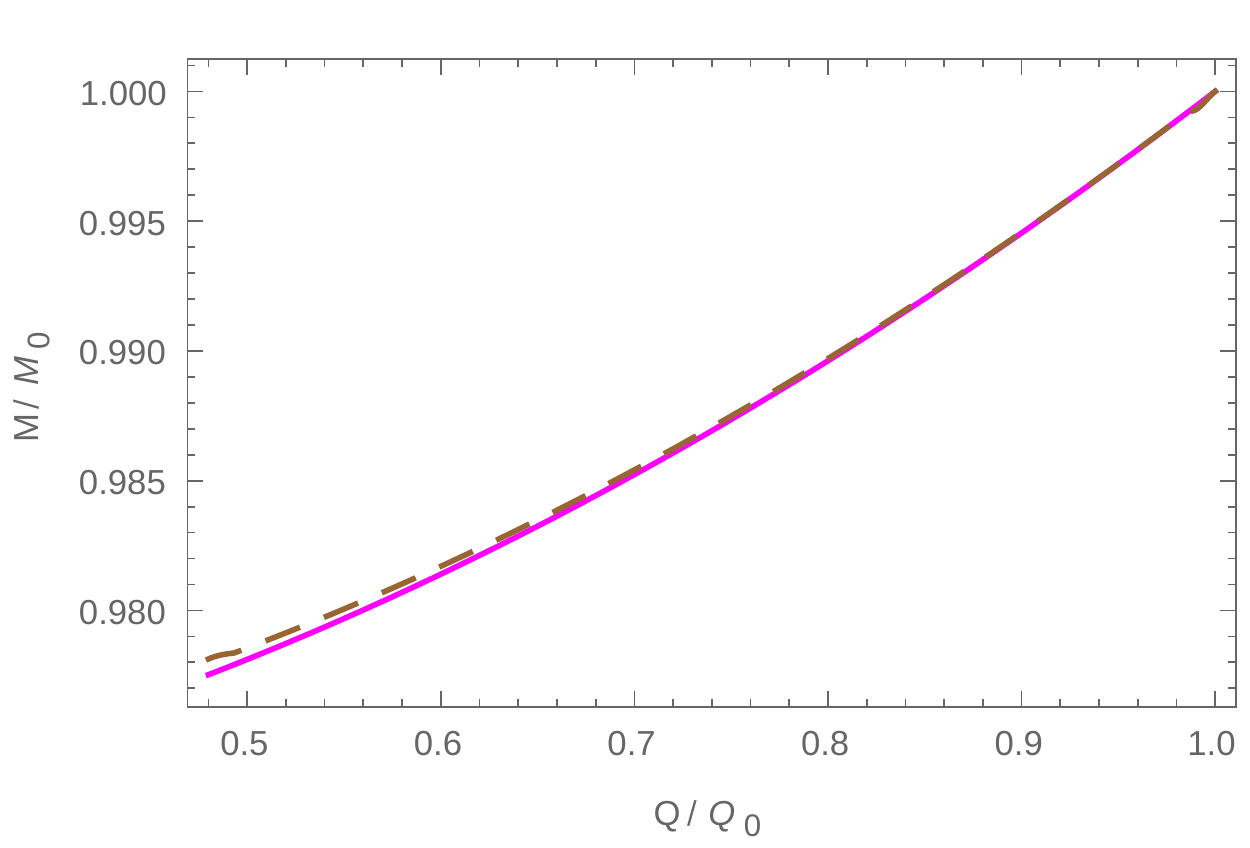}}\hspace{0.5cm}
\subfloat[\label{}]{\includegraphics[width=70mm]{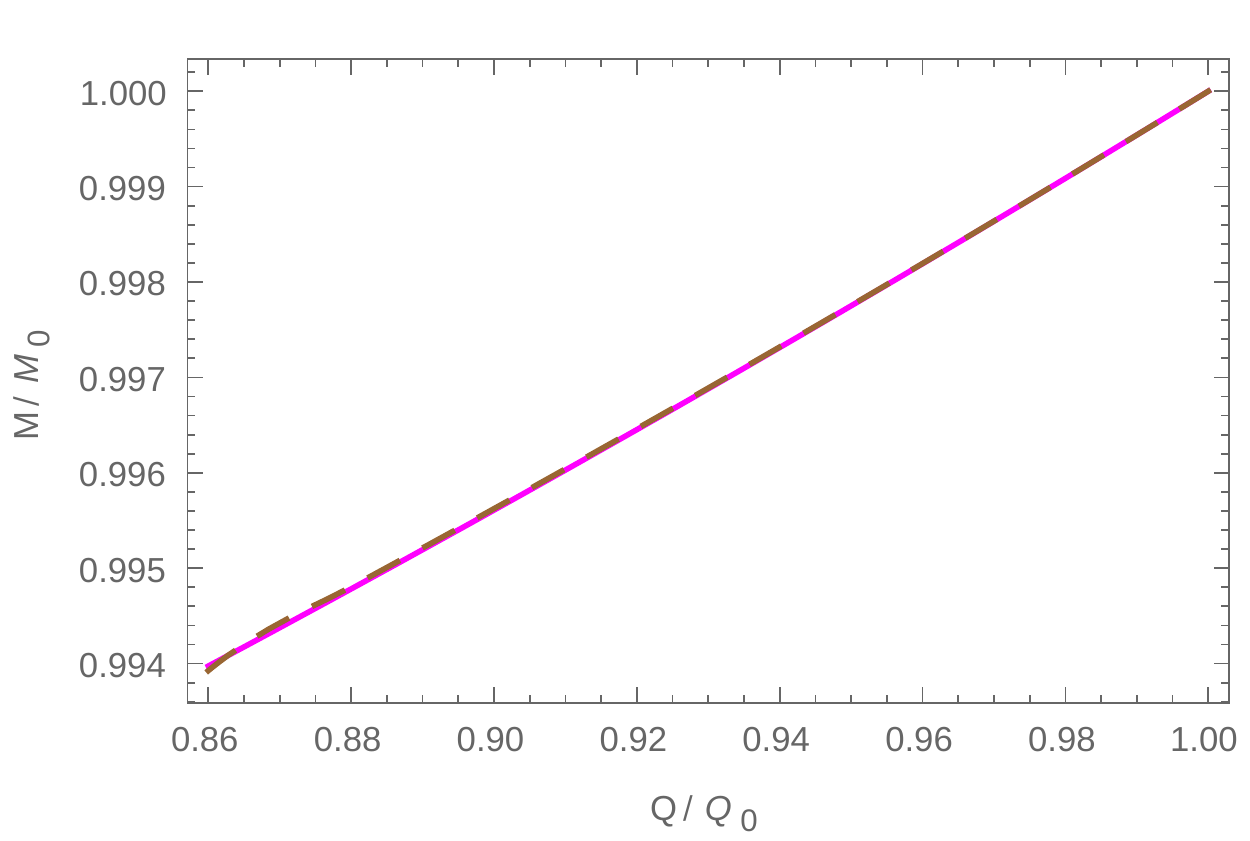}}\\
\subfloat[\label{}]{\includegraphics[width=70mm]{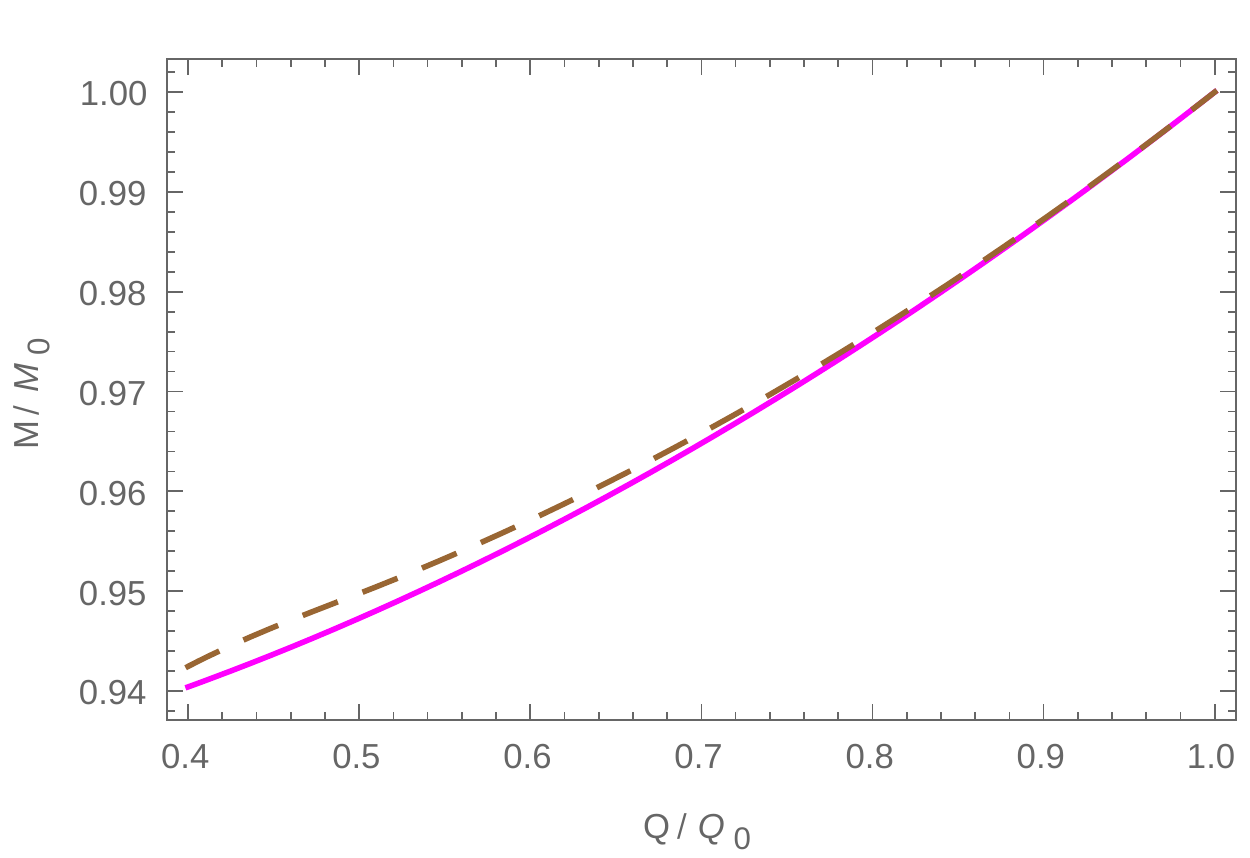}}\hspace{0.5cm} 
\subfloat[\label{}]{\includegraphics[width=70mm]{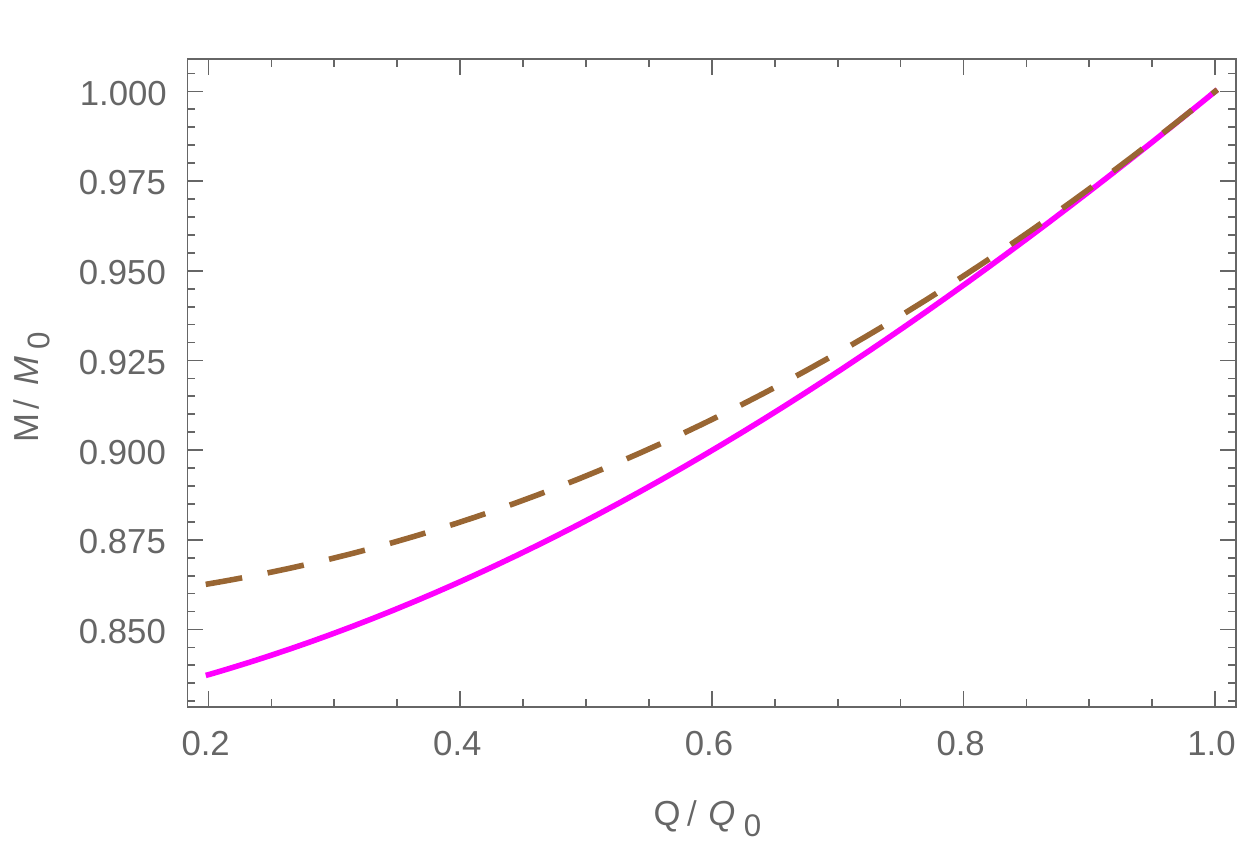}}\vspace{1cm}
\caption{\label{MvsQ}The variation of mass with charge for a supermassive, isolated RN black hole. The solid-line curves correspond to evolution along the geodesic with: \emph{(first 3 subfigures)} $M_*=\SI{3e9}{m}$ and $Q_*$ equal to \textbf{(a)} \num{6e8} \textbf{(b)} \num{8e8} \textbf{(c)} \num{1e9} metres; \emph{(last 3 subfigures)} $M_*=\SI{5e9}{m}$ and $Q_*$ equal to \textbf{(d)} \num{1.5e9} \textbf{(e)} \num{2.5e9} and \textbf{(f)} \num{3.5e9} metres. The dashed curves represent the relation $M(Q)$ obtained by setting the constraints of the respective geodesic as the initial conditions $M_0$ and $Q_0$ used to solve ($\ref{dQdt}$) and ($\ref{dMdt}$)}
\end{figure*}

Now that we have established the nature of the thermodynamic geodesics, the only question yet to be answered is whether a `real' black hole follows a geodesic as it evolves. With this in mind, we solve equations ($\ref{dQdt}$) and ($\ref{dMdt}$) numerically, adopting the values that constrain a particular geodesic as initial conditions\footnote{For instance, in the case of the geodesic with constraints given by Eq. ($\ref{ic}$), we set $Q(t=0) = \SI{8e8}{m}$ and $M(t=0) = \SI{3e9}{m}$}. The result is subsequently used to construct the function $M(Q)$, which is then compared with the solution obtained by rewriting ($\ref{main2}$) in terms of the mass of the black hole and solving the derived equation (while applying the same constraints). Note that here we work with the mass -- rather than the entropy -- because this is the most easily-measured parameter. Additionally, given that a `real' black hole would be discharging, rather than gaining charge -- especially in view of its assumed isolation -- we only consider values of $Q$ that satisfy $Q\leq Q_*$. The results for six of the geodesics are presented in Fig. $\ref{MvsQ}$. 

Provided it is not surrounded by any matter or radiation, a `real' black hole evolves via the emission of thermal radiation and charged-particle pairs. The path it traces in thermodynamic space depends on the initial conditions and would be represented by a curve similar to the dashed ones in Fig. $\ref{MvsQ}$. It can be noted that the smaller values of $Q_*$ ($=Q(0)$) give rise to thermodynamic geodesics which follow the respective dashed curve closely, indicating that the evolution along them coincides with the one stipulated by the combined processes of Hawking radiation and Schwinger pair production. On the other hand, as $Q_*$ increases, the deviation between the solid-line and dashed curves becomes more marked. When this happens, the former lies below the latter, signifying a solution to the geodesic equation which \emph{maximises} changes in $M$ during discharge. Note that since $Q$ would be decreasing, time progresses from right to left in Fig. $\ref{MvsQ}$.  
 
\section{Conclusion}
\label{conclusion}

In this work, using an appropriate thermodynamic metric that emerges from the recently introduced formalism of Geometrothermodynamics \cite{Quevedo2008}, a differential equation is obtained to describe the geodesics in the space of thermodynamic equilibrium states of a supermassive Reissner Nordstr\"{o}m black hole in isolation. The geodesic equation is then solved numerically by considering the processes of Hawking radiation and Schwinger pair production to derive sets of appropriate constraints. We construct our black hole model on the one presented by Hiscock and Weems \cite{Hiscock}. However, we replace their expression for the rate of mass lost due to thermal emission with the Stefan--Boltzmann law for a black body. 

Since we work in the mass representation (i.e. with the mass $M$ of the black hole acting as thermodynamic potential), the space of equilibrium states is coordinatized by the entropy $S$ and electric charge $Q$. Consequently, geodesic curves establish a relation between the two. We propose that this relation extremises changes in the black hole's mass, as can be inferred from the fact that the expression for the entropy of an extremal black hole ($S=\pi Q^2/\hbar$) is an exact solution to the geodesic equation. If such a black hole were to lose charge, the accompanying decrease in mass would be maximum if the black hole remained extremal, i.e. if it evolved along the curve $S=\pi Q^2/\hbar$. On the other hand, in the scenario of an increasing $Q$, variations in $M$ would be minimised if the black hole retained its extremal nature. 

Next, we investigate how the mass varies with charge along a geodesic by solving the geodesic equation in the entropy representation. The results are compared with the actual evolution of the black hole, which can be worked out from equations ($\ref{dQdt}$) and ($\ref{dMdt}$). Whenever a geodesic deviates from the corresponding trajectory mapped out by a `real' black hole in thermodynamic space, it is characterised by a greater loss in mass, implying that -- as was the case with $S=\pi Q^2/\hbar$ -- it maximises changes in $M$. We note that this deviation becomes especially marked as $Q_*$ (the value of $Q$ used to constrain the geodesic equation) approaches its extremal limit. For smaller values of $Q_*$, the proximity of the geodesics to the actual evolution indicates that under certain conditions, the emission of Hawking radiation and charged-particle pairs causes a supermassive, isolated RN black hole to trace a path in thermodynamic space that extremises the thermodynamic length computed from the metric ($\ref{g}$).  

\section*{Acknowledgements}
C. F. would like to thank Prof. H. Quevedo of the Universidad Nacional Aut\'{o}noma de M\'{e}xico, Prof. J. Muscat (Department of Mathematics, University of Malta) and Prof. K. Zarb Adami (Institute of Space Sciences and Astronomy, University of Malta).

\bibliographystyle{ieeetr}
\bibliography{reference.bib}

\end{document}